\documentclass[aps,epsf,twocolumn,showpacs,superscriptaddress,prl]{revtex4}
\usepackage{amsmath}
\usepackage{amssymb}
\usepackage{epsfig}
\usepackage{multirow}
\usepackage{graphicx}
\begin{document}

\title{Fragmentation of fractal random structures}

\author{Eren Metin El\c{c}i}
\email{eren.metin.elci@gmail.com}
\affiliation{Applied Mathematics Research Centre, Coventry
University, Coventry, CV1 5FB, England}
\author{Martin Weigel}
\email{martin.weigel@coventry.ac.uk}
\affiliation{Applied Mathematics Research Centre, Coventry
University, Coventry, CV1 5FB, England}
\affiliation{Institut f\"{u}r Physik, Johannes
Gutenberg-Universit\"{a}t Mainz, Staudinger Weg 7, D-55099 Mainz,
Germany}
\author{Nikolaos G. Fytas}
\email{nikolaos.fytas@coventry.ac.uk}
\affiliation{Applied Mathematics Research Centre, Coventry
University, Coventry, CV1 5FB, England}

\begin{abstract}
  We analyze the fragmentation behavior of random clusters on the lattice under a
  process where bonds between neighboring sites are successively broken. Modeling
  such structures by configurations of a generalized Potts or random-cluster model
  allows us to discuss a wide range of systems with fractal properties including
  trees as well as dense clusters. We present exact results for the densities of
  fragmenting edges and the distribution of fragment sizes for critical clusters in
  two dimensions. Dynamical fragmentation with a size cutoff leads to broad
  distributions of fragment sizes. The resulting power laws are shown to encode
  characteristic fingerprints of the fragmented objects.
\end{abstract}

\pacs{64.60.F-,05.70.Ln,05.50.+q}

\maketitle


Breakup phenomena are ubiquitous in nature and technology \cite{redner:90}. They span
a vast range of time and length scales, including polymer degradation \cite{ziff:86}
as well as collision induced fragmentation of asteroids \cite{housen:90}. In geology,
fragmentation results in the distribution of grain sizes observed in soils; fluids
break up into droplets and fluid structures such as eddies undergo fragmentation
\cite{siebesma:89}. On the subatomic scale, excited atomic nuclei break up into
fragments \cite{berkenbusch:02}. Practical applications, such as mineral processing,
ask for optimizations according to technological requirements and efficiency
considerations \cite{redner:90}. More generally, a wide range of structures from
transport systems to social connections are described by complex networks, whose
degree of resilience against fragmentation is a recent subject of intense scrutiny
\cite{callaway:00,newman:10}.

Considerable effort has been invested in defining and analyzing tractable models of
fragmentation processes \cite{redner:90,astrom:06}. For brittle materials, in
particular, spring or beam models as well as finite-element techniques have been used
to describe the formation and propagation of cracks in problems of fracture and {\em
  instantaneous\/} fragmentation \cite{astrom:06,danku:13,carmona:14}, and these
models allow to describe a range of experimental observations
\cite{oddershede:93,kadono:97,aastrom:04,wittel:04}. In contrast, the fragment-size
distribution (FSD) $n(s,t)$ for {\em continuous\/} fragmentation such as in milling
or the breakup of fluids can be described stochastically by rate equations of the
form \cite{redner:90}
\begin{equation}
  \label{eq:rate_equation}
  \frac{\partial n}{\partial t} = - \int_0^s n(s,t)\,c(s,s',t)\,\mathrm{d}s' 
  + 2 \int_s^\infty n(s',t)\,c(s',s,t)\,\mathrm{d}s',
\end{equation}
where $c(s,s',t) = a(s,t)\,b(s,s',t)$, $a(s,t)$ denotes the fragmentation rate of
clusters of mass $s$, and $b(s,s',t)$ is the conditional probability for an $s$
breakup event to result in a fragment of size $s'$. Here, the first term on the
r.h.s.\ describes the loss of fragments at size $s$ due to breakup, whereas the
second term corresponds to the gain from the breakup of clusters of mass larger than
$s$. In practice, the kernel is normally assumed to be time independent, $c(s,s',t)
\equiv c(s,s')$. Additionally, a description through Eq.~\eqref{eq:rate_equation}
implies a fragmentation process that is spatially homogeneous and independent of
fragment shape --- clearly a drastic simplification. Under such assumptions a useful
scaling theory of solutions can be formulated \cite{cheng:88,krapivsky:94}.

Much less progress has been made in terms of results beyond this mean-field
approximation. What is the relation between geometrical properties of fragmented
objects and the resulting FSDs? This has been studied for loop-less structures such
as intervals \cite{manna:92,krapivsky:00,krapivsky:00a,dean:02} and trees
\cite{kalay:14}. For higher-dimensional shapes the only results to date concern the
fragmentation of percolation clusters \cite{gyure:92,debierre:97,cheon:99,lee:14}. It
was demonstrated numerically there that the fragmentation rate $a(s)$ as well as the
conditional breakup probability $b(s,s')$ exhibit power-law scaling.

In the present Letter we discuss fragmentation within a generalization of the
percolation model with bond activation probability $p$, additional cluster weight
$q$, and partition function
\begin{equation}
  Z_\mathrm{RC} = \sum_{{\cal G}' \subseteq {\cal G}}  p^{b({\cal G}')} (1-p)^{{\cal
      E}-b({\cal G}')}q^{k({\cal
      G}')},\;\;p, q > 0,
  \label{eq:cluster_weight}
\end{equation}
known as the random-cluster (RC) model \cite{grimmett:book}. Here, $b({\cal G}')$
denotes the number of active edges out of a total number ${\cal E}$ of edges of
${\cal G}$, and $k({\cal G}')$ is the resulting number of connected components in the
spanning subgraph ${\cal G}' \subseteq {\cal G}$. Variation of $q$ allows the model to
describe a wide range of fractal structures and different connectivities
\cite{deng:10}, thus accounting for the differences in mechanical response of a range
of materials \cite{stauffer:book,broedersz:11}. The model includes as particular
limits percolation ($q\to 1$) and the Ising model ($q=2$). As $p$ is increased, a
giant or percolating cluster appears in the system. For sufficiently large $q$, this
transition is of first order, while for small $q$ it is continuous. For the square
lattice the transition occurs at coupling $p_c = \sqrt{q}/(q+\sqrt{q})$, being
continuous for $q \le 4$ \cite{wu:82a}.

\begin{figure}[tb]
\includegraphics*{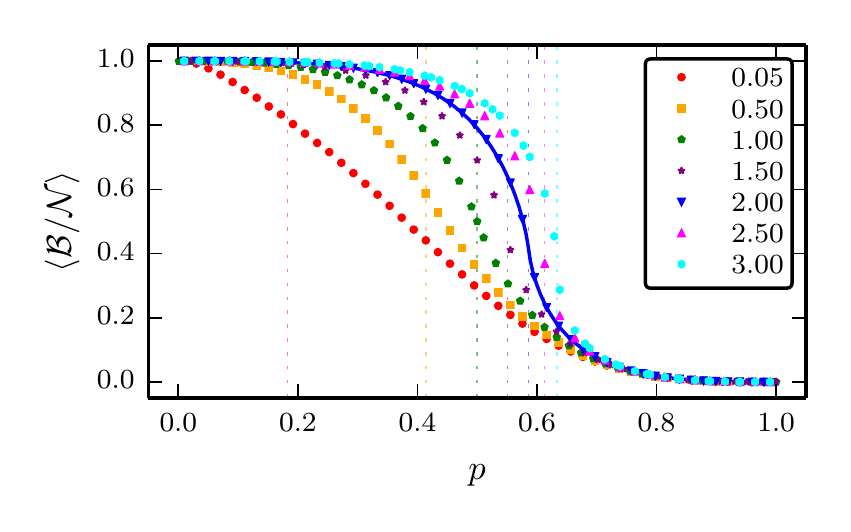}
\caption{\label{fig:bridges0} Proportion of bridges among active edges in the
  equilibrium random-cluster model for different values of $q$. Simulation data are
  for systems sizes $L= 64$ ($q\ne 1$) and $L=2048$ ($q=1$), respectively. The solid
  line denotes the exact result for $q=2$ and $L\to \infty$. The vertical dashed
  lines specify the location of the critical point. Simulations were performed using
  the algorithm described in Ref.~\cite{elci:13}.}
\end{figure}

The fragmentation processes discussed here start from an equilibrium configuration of
the RC model \eqref{eq:cluster_weight} with bond weight $p$. The removal of a
randomly chosen bond can result in a breakup, creating an additional fragment. In
this case, the bond is called a {\em bridge\/}. Such a consumption of bridges can
serve as a model for the degradation of porous material such as in the combustion of
charcoal particles \cite{huang:96}.  Similarly, it may describe the breaking of
chemical bonds in polymers. The structural resilience under bond removal then depends
on the density $\mathcal{B}$ of bridges among all active bonds $\mathcal{N}$. Figure
\ref{fig:bridges0} shows $\langle\mathcal{B}/\mathcal{N}\rangle$ for the equilibrium
square-lattice RC model. Incidentally, it is seen that the change of the relative
bridge density and hence the change in fragility of the configuration becomes maximal
at the critical coupling $p_c$. This is when a significant fraction of fragmentation
events first appears, an effect connected to the (self-)entanglement of critical
clusters \cite{elci:prep}. In particular, as will be shown below, the behavior of
$\langle {\cal B}/{\cal N}\rangle$ near $p_c$ is governed by the specific-heat
exponent $\alpha$, which implies a divergent slope for $q\geq 2$.

\begin{figure}
\includegraphics*{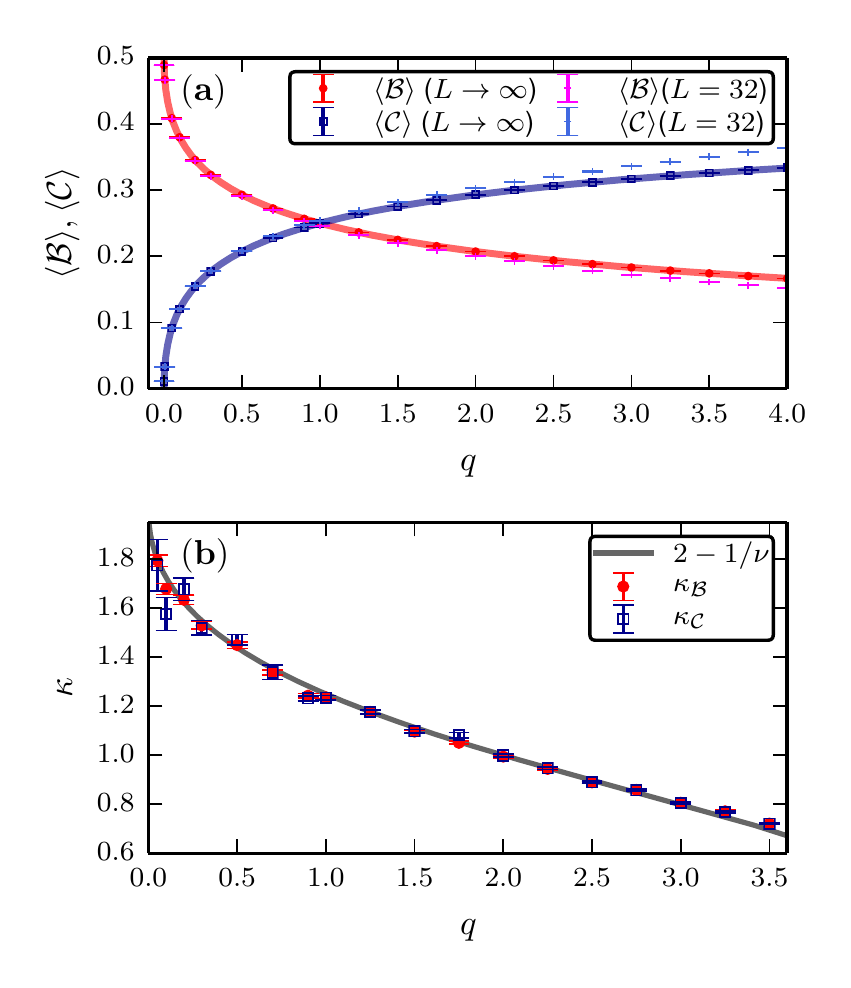}
\caption{\label{fig:bridges}(a) Asymptotic critical density of bridges, $\langle {\cal
    B}\rangle$, and non-bridges, $\langle {\cal C} \rangle$. 
    (b) Finite-size correction exponent $\kappa$ for the bridge density according to
  Eq.~\eqref{eq:kappa} for the random-cluster model on the square lattice.  }
\end{figure}

Let us first discuss what happens for a single bond removal if we start at the
critical point $p = p_c$ at time $t=0$. What is the form of $a(s)\,b(s,s') \equiv
a(s,0)\,b(s,s',0)$ for this case? A standard ansatz for Eq.~\eqref{eq:rate_equation}
is $a(s) \sim s^\lambda$, where a range $\lambda \le 1$ of values is found in
experiments \cite{redner:90}. A {\em shattering transition\/} occurs for $\lambda \to
0$ \cite{mott:47,mcgrady:87}. To determine $\lambda$ for the critical RC model,
consider the total number of bridges,
\begin{equation}
    \frac{\sum_s s\,n(s,0)a(s)}{\sum_s s\,n(s,0)}\sim \int s^{-\tau +1 +\lambda}e^{-cs}\,\mathrm{d}s \sim
  L^{(\tau - \lambda)/(\sigma \nu)},
  \label{eq:total_bridges}
\end{equation}
where we have used the scaling form of the critical FSD, $n(s,0)\sim
s^{-\tau}e^{-cs}$ as well as the relations $c\sim \vert
p-p_c\vert^{\frac{1}{\sigma}}$ and $\vert p-p_c\vert \sim L^{-\frac{1}{\nu}}$, where
$L$ is the linear dimension, and $\nu$, $\sigma$, $\tau$ are standard critical
exponents \cite{stauffer:book}.  From Fig.~\ref{fig:bridges0} it appears that the
density of bridges is asymptotically non-vanishing. This is seen more clearly in our
results for the critical bridge density shown in Fig.~\ref{fig:bridges}(a). Hence the
average number of bridges in \eqref{eq:total_bridges} must grow as $L^d$, implying $d
= (\tau-\lambda)/\sigma\nu$.  With the exponent identities $\sigma\nu = 1/d_F$ and
$\tau = 1 + d/d_F$, where $d_F$ is the critical cluster fractal dimension, this shows
that
\begin{equation}
  \label{eq:lambda}
  \lambda = 1,  
\end{equation}
independent of $q$. Hence the breakup is spatially homogeneous. This confirms
previous numerical results for $q\to 1$ \cite{edwards:92,cheon:99}.

While Eq.~\eqref{eq:lambda} rests on the numerical observation of
Fig.~\ref{fig:bridges} for the square lattice, it is more general. By applying a
rigorous analysis of the \emph{influence} of an edge \cite{grimmett:book}, we can
express the $p$-derivative of the corresponding partition function $Z_{\rm RC}$ in
terms of $\langle {\cal B} \rangle$ and equate this expression with the standard
result, identifying the $p$-derivative of $Z_{\rm RC}$ with $\langle {\cal N}
\rangle$ \cite{elci:prep}. We deduce that for the RC model on an arbitrary graph the
bridge and bond densities, $\langle {\cal B}\rangle $ and $\langle {\cal N}\rangle$,
are related as
\begin{equation}
  \label{eq:bridges_and_edges}
  \langle {\cal B}\rangle = \frac{\langle {\cal N}\rangle - p}{(1-p)(1-q)}
\end{equation}
such that, in general, the bridge density is non-vanishing whenever the edge density
is positive. The singular case $\langle {\cal N}\rangle = p$ corresponds to the
percolation limit $q\to 1$, for which a closer analysis shows that $\langle {\cal
  B}\rangle$ still is finite. Hence \eqref{eq:lambda} holds for the RC model on any
graph for any bond probability $0 < p < 1$, on or off criticality. For the square
lattice, the critical edge density is $\langle {\cal N}\rangle_c = 1/2$
\cite{wu:82a}, such that we find the exact expression
\begin{equation}
\langle {\cal B}\rangle_c = \frac{1}{2}\frac{1}{1+\sqrt{q}},
\label{eq:asymptotic_bridges}
\end{equation}
generalizing a recent result for percolation \cite{xu:14}.
Figure ~\ref{fig:bridges}(a) shows our simulation data together with the asymptotic
result \eqref{eq:asymptotic_bridges}.  Relation \eqref{eq:bridges_and_edges} shows
that the finite-size corrections to $\langle{\cal B}\rangle$ are given by the
corrections to the edge density $\langle {\cal N} \rangle$ , which in turn is related
to the energy density of the Potts model $u = -2\langle {\cal N}\rangle/p$
\cite{grimmett:book}. Standard scaling arguments \cite{privman:privman} lead to
\begin{equation}
  \label{eq:kappa}
  u_L = u_\infty + A_u L^{-\kappa} + o(L^{-\kappa}),  
\end{equation}
where $\kappa = (1-\alpha)/\nu = d-1/\nu$, in agreement with our data for the
finite-size corrections to the density of bridges shown in
Fig.~\ref{fig:bridges}(b). As a consequence of \eqref{eq:bridges_and_edges} one can
show that the $p$-derivative of $\langle {\cal B} \rangle (p)$ has a power-law
singularity at the critical point $p_c$. This is governed by the specific-heat
exponent $\alpha$. Similar results can be derived for the density $\langle {\cal C}
\rangle$ of non-bridges \cite{elci:prep}.


\begin{figure}
    \includegraphics*{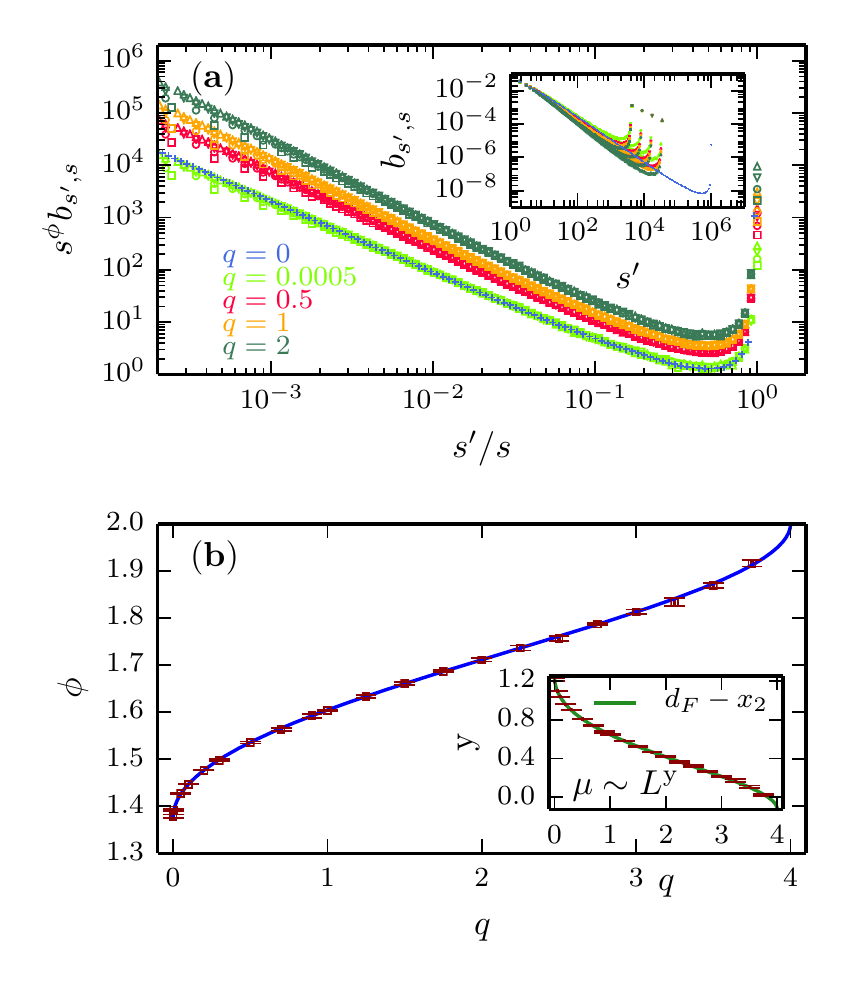}
    \caption{\label{fig:phi} (a) Re-scaled conditional fragmentation probability
      $b(s',s)$ according to Eq.~\eqref{eq:fss_daughter} for different values of the
      cluster coupling $q$. (b) Scaling exponent $\phi$ of daughter
      clusters in the fragmentation of the square-lattice RC model as compared to the
      exact result \eqref{eq:phi_exp}. The inset shows the scaling
      exponent of the ensemble average daughter cluster size $\langle s'\rangle$.}
\end{figure}

Cluster breakup rates are hence proportional to the cluster size. The typical size of
fragments created in a breakup at criticality is encoded in the probability
$b(s,s')$. The scale-free nature of the critical RC model suggests a large-$s$
scaling form
\begin{equation}
b_{s',s} \sim s^{-\phi} \mathcal{G}\left(\frac{s'}{s},\frac{s}{L^{d_F}}\right), 
\label{eq:fss_daughter}
\end{equation}
which is compatible with exact results for percolation in 1D and on the Bethe lattice
\cite{gyure:92}. To relate $\phi$ to previously established critical exponents, we
multiply Eq.~\eqref{eq:fss_daughter} by $s'$ and then integrate to find that $\mu_s
\sim s^{2-\phi} {\cal H}(s/L^{d_F})$. Using a finite-size scaling form of the overall
FSD \cite{deng:10} we conclude that the scaling of the ensemble average daughter
cluster size is $\langle s'\rangle \sim L^{d_F(3-d/d_F-\phi)}$.  On the other hand,
one can show \cite{elci:prep} that this is proportional to the average of
$C_{\min,2}$, the size of the smaller of the two clusters attached to two neighboring
disconnected vertices \cite{deng:10,elci:13}. In Ref.~\cite{deng:10} it was shown
that $\langle C_{\min,2}\rangle \sim L^{d_F-x_2}$, where $x_2$ is known as two-arm
exponent, which implies
\begin{equation}
\phi = 2+(x_2-d)/d_F = 2 - d_R/d_F,
\label{eq:phi_exp}
\end{equation}
where $d_R = d-x_2$ is the red-bond fractal dimension, and $d$ denotes the spatial
dimension. Again, this confirms and generalizes previous results for bond percolation
\cite{edwards:92,debierre:97}. Another special case concerns the uniform spanning
tree ensemble $p,q\rightarrow 0$ with $q/p\rightarrow 0$ for which $\phi \rightarrow
\frac{11}{8}$, in agreement with Ref.~\cite{manna:92}. As the data in
Fig.~\ref{fig:phi}(b) show, our numerical simulations for the full range $0\le q\le
4$ are in perfect agreement with Eq.~\eqref{eq:phi_exp}. More generally,
Fig.~\ref{fig:phi}(a) demonstrates the validity of the scaling form of
Eq.~\eqref{eq:fss_daughter}, showing an excellent collapse of data for different
cluster and system sizes onto scaling functions parametrized by $q$. Notably, in
contrast to recent claims in Ref.~\cite{schroder:13}, for the RC model clusters do
not typically break up into equally sized fragments.

\begin{figure}
  \includegraphics*{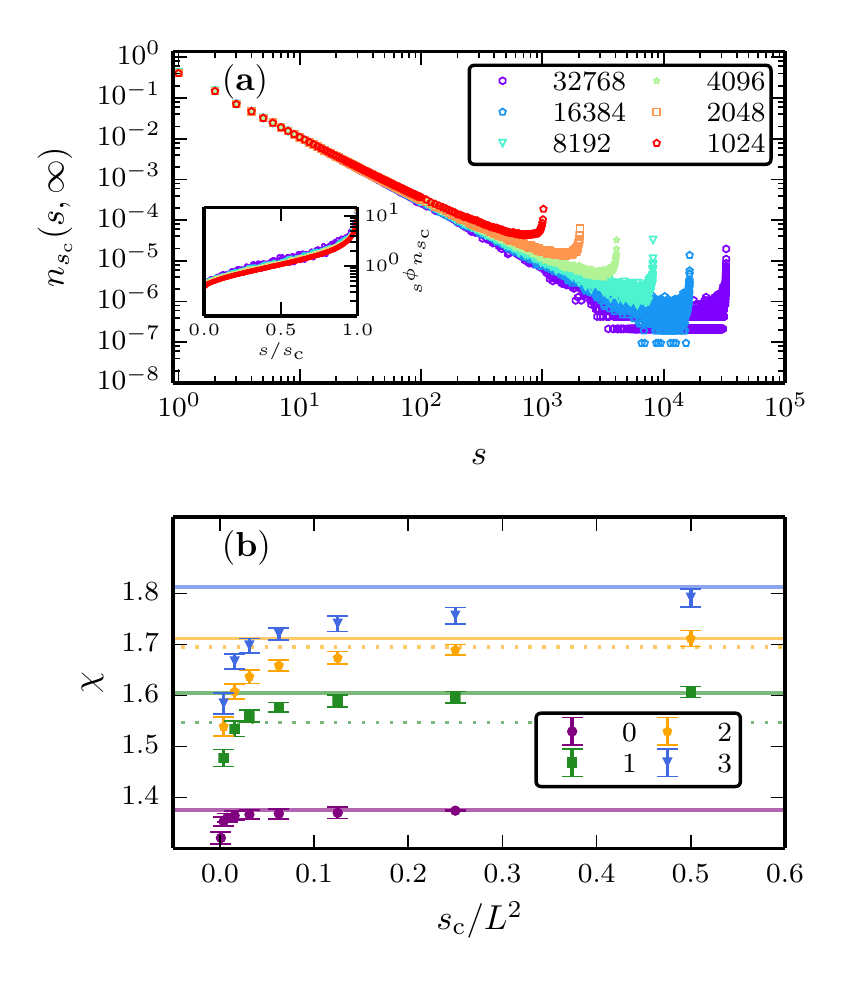}
  \caption{\label{fig:kinetics} (a) Final FSD from iterative
    fragmentation of the critical giant component for $q\to 1$ and different values
    of the cutoff $s_\mathrm{c}$ ($L=256$). (b) Dynamical fragmentation exponent
    $\chi$ according to Eq.~\eqref{eq:kinetic_sizes} for critical initial
    configurations of cluster weight $q$. The solid horizontal lines mark the exact
    values of $\phi$ in 2D and the dotted lines show estimates of $\phi$ for 3D
    \cite{deng:07}.  }
\end{figure}

We now generalize to the case of dynamic or continuous fragmentation processes,
corresponding to the sequential removal of bonds, or $t > 0$. In general, we must
then expect the equilibrium description to break down and $c(s,s',t)$ to be time
dependent. Random bond removal drives any initial configuration into a stationary
state where all fragments only consist of one vertex. For real fragmentation
processes, however, one rather expects a critical particle size $s_\mathrm{c}$ below
which there is no further breakup \cite{redner:90}. This could come about, for
instance, through surface tension for the breakup of droplets, via the chosen
geometry in a mill, or through energetic limitations in nuclear fragmentation events.
Limited fragmentation has been studied for simpler geometries such as intervals and
trees \cite{krapivsky:00,krapivsky:00a,dean:02,majumdar:02} (see also
Ref.~\cite{huang:96}). For fragmentation processes again starting from critical
equilibrium configurations (here for $q\to 1$), the final FSD below the cutoff $s_c$
is shown in Fig.~\ref{fig:kinetics}(a). Over a range of fragment sizes increasing
with $s_c$, the data clearly follow a power-law. Additionally, the dependence on
$s_c$ is only via the ratio $s/s_c$, resulting in a scaling form
\begin{equation}
    n_{s_c}(s, \infty) \sim s^{-\chi} {\cal F}\left(\frac{s}{s_\mathrm{c}}\right),
    \label{eq:kinetic_sizes} 
\end{equation}
with a dynamic fragmentation exponent $\chi$. Figure~\ref{fig:kinetics}(b) summarizes
the result of power-law fits to the decay displayed in Fig.~\ref{fig:kinetics}(a) for
different cluster weights $q$. For sufficiently large cutoffs, we find that $\chi$
coincides with the exponent $\phi = 2 - d_R/d_F$ characteristic of {\em
  equilibrium\/} fragmentation. The deviations for $s_c/L^2 \ll 1$ are an effect of
the scaling function ${\cal F}$ of Eq.~\eqref{eq:kinetic_sizes}. Moreover, not only
the power-law decay but the full scaling form \eqref{eq:kinetic_sizes} of the final
FSD is fully supported by our data, as is illustrated in the scaling collapse shown
in the inset of Fig.~\ref{fig:kinetics}(a).

While the close relation of dynamical fragmentation with critical equilibrium
properties is at first surprising, it can be understood from the nature of the
breakup process. Due to the shape of the breakup kernel shown in Fig.~\ref{fig:phi},
the process is dominated by ``abrasive'' breakup, i.e., small daughter
clusters. Representing the fragmentation events in a genealogical tree, we indeed
typically find one long branch, related to the erosion of the giant component, with
sub-branches of only a few steps \cite{elci:prep}. In contrast, uniform breakups
would result in a statistically balanced genealogical tree \cite{dean:02}. We hence
find the basic assumption in the mean-field model \eqref{eq:rate_equation} of taking
the breakup kernel $c(s,s')$ to be independent of time to be rather appropriate for
the model studied here.

In summary, we have first given a scaling description of the fragmentation of
critical configurations in the RC model. The density of fragmenting edges is
independent of cluster size, implying $\lambda = 1$. The daughter-size function
assumes a scaling form with a scaling index connected to the two-arm
exponent. Further conclusions follow from the general result
\eqref{eq:bridges_and_edges} \cite{elci:prep}. Investigating the asymptotic FSD under
continuous fragmentation with a cutoff $s_\mathrm{c}$, we find that this
non-equilibrium process is determined by the equilibrium critical behavior with a
final FSD described by the equilibrium exponent $\phi$. The FSD hence reveals
structural characteristics of the initially fragmented object. The insensitivity to microscopic
details implied by the universality of critical phenomena indicates that our results
for dynamic fragmentation should be comparable also to experiments. In fact, the size
exponents found experimentally span a range of around $1.2$ to $1.9$
\cite{redner:90,astrom:06,timar:10} which is also covered by our model on varying
$q$, cf.\ Fig.~\ref{fig:kinetics}.

We have restricted ourselves to the case of bond fragmentation. A more general
situation occurs for the deletion of vertices producing up to $z$ fragments, where
$z$ is the coordination number of the lattice. In this case we find that the binary
branch is still strongly dominant. Preliminary investigations indicate a connection
between the statistics of such breakup events and generalizations of
Eq.~\eqref{eq:fss_daughter}, where the scaling exponents $\phi^{(k)} = 2-(d-x_k)/d_F$ of
breakups with $k$ fragments are governed by the corresponding multi-arm exponents
$x_k$ \cite{deng:10}.

Our results also carry over to lattices in 3D. In fact, we have studied the
fragmentation of clusters of Eq.~\eqref{eq:cluster_weight} on the simple cubic
lattice and confirmed that $\lambda = 1$. Selected results for the value of $\phi$ in
3D also shown in Fig.~\ref{fig:kinetics}(b) indicate that a very similar range of
FSDs can be described there. For the dynamical fragmentation process, we find that
fragmenting solid instead of fractal objects also leads to algebraically decaying
FSDs, however governed by a different set of exponents \cite{elci:prep}. Beyond the
implications of the present work for fragmentation processes in nature and industry,
an exciting extension concerns the fragmentation of random graphs and networks in
order to model resilience.

\begin{acknowledgments}
  We thank Timothy M. Garoni and Youjin Deng for valuable discussions. We acknowledge
  funding from the DFG (WE4425/1-1) and the EC FP7 Programme (PIRSES-GA-2013-612707).
\end{acknowledgments}








\end{document}